\documentclass[%
superscriptaddress,
preprint,
 amsmath,amssymb,
 aps,
prd,
floatfix,
]{revtex4-1}

\newcommand{\bA}{\boldsymbol{A}}
\newcommand{\bt}{\boldsymbol{t}}

\newcommand{\I}{\mathrm{i}}
\usepackage{tikz}
\usetikzlibrary{decorations.pathreplacing}
 \usetikzlibrary{plotmarks}
\usepackage{subfigure}
\usepackage{booktabs}
\usepackage{graphicx}
\usepackage{dcolumn}
\usepackage{bm}
\usepackage{youngtab}
\usepackage{ytableau}

\usepackage{braket}
\usepackage{multirow}
\usepackage{dsfont}
\usepackage{mathtools}
\usepackage{xcolor}
\usepackage[normalem]{ulem}

\begin{document}

\newcommand{\vsub}[1]{%
  \begin{smallmatrix}#1\end{smallmatrix}%
}
\newcommand{\cbl}[1]{\textcolor{red}{[[[CBL: #1]]]}}
\newcommand{\cma}[1]{\textcolor{red}{[[[CMA: #1]]]}}

\title{Chiral-spin symmetry emergence in baryons and eigenmodes of the Dirac operator}

\author{M.~Catillo}
\author{L.~Ya.~Glozman}
\author{C.~B.~Lang}

\affiliation{Institute of Physics, University of Graz, 8010 Graz, Austria}
 
\date{\today}

\begin{abstract}
Truncating  the low-lying modes
of the  lattice Dirac operator results in an emergence of
the chiral-spin symmetry $SU(2)_{CS}$ and its flavor extension
$SU(2N_F)$  in hadrons. These are symmetries of 
the  quark - chromo-electric interaction and include
chiral symmetries as subgroups. Hence the quark - chromo-magnetic
interaction, which breaks both symmetries, is located at least predominantly in the
near - zero modes.
Using as a tool the expansion of propagators
into eigenmodes of the Dirac operator we here  analytically study
effects of a gap in the eigenmode spectrum on baryon correlators. We find 
that  both $U(1)_A$ and $SU(2)_L \times SU(2)_R$ emerge automatically
if there is a gap around zero. Emergence of larger  $SU(2)_{CS}$
and $SU(4)$ symmetries requires in addition a microscopical dynamical
input about   the higher-lying modes and their symmetry structure.
\end{abstract}
\maketitle


\section{Introduction}

In a number of lattice spectroscopical studies with a
chirally-invariant Dirac operator upon artificial truncation
of the lowest modes of the Dirac operator \cite{ls,gls} a large degeneracy was
discovered in mesons \cite{Denissenya:2014poa,Denissenya:2014ywa,Denissenya:2015mqa}
and baryons \cite{Denissenya:2015woa}.
 Corresponding symmetry groups, $SU(2)_{CS}$ and  $SU(2N_F)$ \cite{Glozman:2014mka,Glozman:2015qva}, turned out to be larger
than the chiral symmetry $SU(N_F)_L \times SU(N_F)_R \times U(1)_A$
of the QCD Lagrangian. The chiral-spin symmetry group $SU(2)_{CS}$ has 
$U(1)_A$ as a subgroup 
while its flavor extension $SU(2N_F)$ contains both $SU(N_F)_L \times SU(N_F)_R \times U(1)_A$ and $SU(2)_{CS}$ as subgroups. The chiral-spin transformations from  $SU(2)_{CS}$ includes rotations 
that mix the left- and
right-handed components of the quark field. Obviously these symmetries are
not  symmetries of a free Dirac equation or of the QCD Lagrangian. However,
they are symmetries  of the Lorentz-invariant fermion charge operator and (in a given reference frame) of the quark - chromo-electric interaction
while the interaction of quarks with the chromo-magnetic field and the quark
kinetic term break them. Consequently the emergence of $SU(2)_{CS}$ and  $SU(2N_F)$ upon truncation of the low-lying modes tells that while the confining quark - electric
interaction is distributed among all modes of the Dirac operator,
the quark - magnetic interaction is located at least predominantly
in the near - zero modes.
Some unknown microscopic dynamics should be responsible for this
phenomenon.

These symmetries emerge naturally, i.e. without any explicit truncation,   
in hot QCD above the pseudocritical temperature \cite{R1,R2,G2}, where the  near-zero
modes of the Dirac operator are suppressed by temperature \cite{Tomiya:2016jwr}. Consequently elementary objects in that range
are not  free quarks and gluons but
rather chirally symmetric quarks  bound   by the chromo-electric
field into  color singlet objects, like a "string".

According to the Banks-Casher relation \cite{BC} the chiral symmetry
breaking quark condensate is proportional to the density of the near-zero modes.
A gap
in the low lying Dirac eigenmode spectrum induces restoration of 
$SU(N_F)_L \times SU(N_F)_R$ symmetry.  It was shown that it also induces
restoration of $U(1)_A$ in the $J=0$ mesons \cite{Cohen}.
Analytical study of the $J=0$ and $J=1$ isovector meson propagators in terms of the
eigenmodes of the Dirac operator revealed that all meson correlators that are
connected by the $U(1)_A$ and/or $SU(2)_L \times SU(2)_R$ transformations
get necessarily degenerate if such a gap exists in the Dirac spectrum \cite{Lang}.
However, a possible emergence of $SU(2)_{CS}$ and of $SU(2N_F)$ requires further
dynamical properties encoded in certain matrix elements. Here we extend this analysis to baryons and show that the same conclusions remain valid in this case as well.

\section{Chiral-spin symmetry }

The  $SU(2)_{CS}$ chiral-spin transformations for quarks are given by 
\begin{equation}
\label{V-def}
  \psi \rightarrow  \psi^\prime = \exp \left(i  \frac{\varepsilon^n \Sigma^n}{2}\right) \psi  \; ,
\end{equation}where generators,
defined in the Dirac spinor space are 
\begin{equation}
 \bf {\Sigma} = \{\gamma_k,-\I \gamma_5\gamma_k,\gamma_5\}\,.
\label{eq:su2cs_}
\end{equation}
Here $\gamma_k$, $k=1,2,3,4$, are hermitian Euclidean gamma-matrices, obeying the 
anticommutation relations
\begin{equation}
\gamma_i\gamma_j + \gamma_j \gamma_i =
2\delta_{ij}; \qquad \gamma_5 = \gamma_1\gamma_2\gamma_3\gamma_4\,.
\label{eq:diracalgebra}
\end{equation}
Different $k$ define  
four-dimensional representations that can be reduced into two-dimensional irreducible ones. 
The $\mathfrak{su}(2)$ algebra
\begin{equation}
[\Sigma^a,\Sigma^b]=2\I\epsilon^{abc}\Sigma^c
\end{equation}
is satisfied for any $k$ in Eq. (\ref{eq:su2cs_}).

$U(1)_A$ is a subgroup of $SU(2)_{CS}$.
The  $SU(2)_{CS}$ transformations mix the
left- and right-handed fermions and different representations of the
Lorentz group. The free massless quark Lagrangian  and Dirac equation
do not have this symmetry.

Extending  the direct product $SU(2)_{CS} \times SU(N_F)$
one obtains an $SU(2N_F)$ group. The chiral
symmetry group of QCD $SU(N_F)_L \times SU(N_F)_R \times U(1)_A$ is a subgroup
of $SU(2N_F)$.
The $SU(2N_F)$ transformations  are given by
\begin{equation}
\label{W-def}
\psi \rightarrow  \psi^\prime = \exp\left(\frac{\I}{2}\,\epsilon^m T^m\right) \psi\; ,
\end{equation}
where $m=1,2,...,(2N_F)^2-1$. The set of $(2N_F)^2-1$ generators is
\begin{align}
T^m =\{
(\tau^a \otimes \mathds{1}_D),
(\mathds{1}_F \otimes \Sigma^n),
(\tau^a \otimes \Sigma^n)
\},
\end{align}
with  the flavor generators $\tau$ with flavor index $a$ and 
$n=1,2,3$ is the $SU(2)_{CS}$ index.

The fundamental
vector of $SU(2N_F)$ at $N_F=2$ is
\begin{equation}
\psi =\begin{pmatrix} u_{\textsc{R}} \\ u_{\textsc{L}}  \\ d_{\textsc{R}}  \\ d_{\textsc{L}} \end{pmatrix}. 
\label{eq:quarkvec}
\end{equation}
The $SU(2)_{CS}$ and $SU(2N_F)$ groups are not symmetries
of the QCD Lagrangian as a whole.

In a given reference frame the quark-gluon interaction Lagrangian in Minkowski space
can be splitted into temporal and spatial parts:
\begin{equation}
\overline{\psi}   \gamma^{\mu} D_{\mu} \psi = \overline{\psi}   \gamma^0 D_0  \psi 
  + \overline{\psi}   \gamma^i D_i  \psi .
\label{cl}
\end{equation}
\noindent
Here $D_{\mu}$ is a covariant derivative that includes
interaction of the quark field $\psi$ with the  gluon field $\bA_\mu$,
\begin{equation}
D_{\mu}\psi =( \partial_\mu - ig \frac{\bt \cdot \bA_\mu}{2})\psi.
\end{equation}
The temporal term includes an interaction of the color-octet
 charge density 
\begin{equation}
\bar \psi (x)  \gamma^0  \frac{\bt}{2} \psi(x) = \psi (x)^\dagger  \frac{\bt}{2} \psi(x)
\label{den}
\end{equation}
with the electric  
part of the gluonic gauge field. 
It is invariant  under any unitary transformation acting in the Dirac and/or
flavor spaces. In particular it is a singlet under
 $SU(2)_{CS}$  and  $SU(2N_F)$ groups. 
 The spatial part consists of  a quark kinetic term
and   interaction with the magnetic part of the gauge field.  It breaks 
 $SU(2)_{CS}$ and $SU(2N_F)$.   We conclude that  interaction
 of electric and magnetic components
 of the gauge field with fermions can be distinguished
 by symmetry. 

In order to discuss the notions "electric" and "magnetic"
one needs to fix the reference frame. An invariant mass of the hadron
is  the rest frame energy. Consequently, to discuss physics
of hadron mass generation it is natural to use the hadron rest frame.

In refs. \cite{Denissenya:2014poa,Denissenya:2014ywa,Denissenya:2015mqa}
and  \cite{Denissenya:2015woa}
meson and baryon masses have been  extracted from the asymptotic slope of the
 rest frame $t$-direction Euclidean correlator
 \begin{equation}
C_\Gamma(t) = \sum\limits_{x, y, z}
\braket{\mathcal{O}_\Gamma(x,y,z,t)
\mathcal{O}_\Gamma(\mathbf{0},0)^\dagger},
\label{eq:momentumprojection}
\end{equation}
where $\mathcal{O}_\Gamma(x,y,z,t)$ is an operator that creates a quark-antiquark pair for mesons  or three quarks for baryons
with fixed quantum numbers. Truncation of the near-zero modes of
the Dirac operator resulted in emergence of the $SU(2)_{CS}$ and $SU(2N_F)$
symmetries in hadrons.

This implies that a confining $SU(2)_{CS}$- and $SU(2N_F)$-symmetric
quark-electric interaction is distributed among all modes of the
Dirac operator. At the same time the quark-magnetic interaction, that breaks both symmetries, is located only in the low-lying modes. Consequently 
truncating the low-lying modes results in emergence of symmetries
in the spectrum of hadrons.

\section{Chiral and chiral-spin transformations of nucleon operators}

In Ref. \cite{Denissenya:2015woa} the emergence of the $SU(2)_{CS}$
symmetry in nucleons upon truncation of the lowest-lying
modes of the Dirac operator was studied on the lattice. In particular it was demonstrated that 
{\em correlators} along the time direction calculated with different
nucleon operators
that are not connected by chiral $U(1)_A$ and/or $SU(2)_L \times SU(2)_R$
transformations but connected by the chiral-spin transformation
(\ref{V-def})-(\ref{eq:su2cs_}) with $k=4$ get degenerate. As discussed
in the introduction our main objective here is to analyse
which conditions  would be sufficient for emergence of chiral and chiral-spin
symmetries in nucleons upon the low-mode truncation (or suppression). To this end we first
classify the nucleon operators with respect to chiral and chiral-spin
transformations. Such a classification of nucleon operators (with spin zero
diquark) for $U(1)_A$,
 $SU(2)_L \times SU(2)_R$ and $SU(2)_{CS}, k=4$ transformations
 is discussed below.

A complete set of nucleon operators ($J=1/2,I=1/2, P=\pm 1$) with spin-zero diquarks consists of four operators \cite{CG} of the following form:
\begin{equation}
N_{\pm}^{(i)} = \epsilon_{abc}\mathcal{P}_{\pm}\Gamma_1^{(i)}u_a\{d_b^T \Gamma_2^{(i)}u_c - u_b^T \Gamma_2^{(i)}d_c\},
\label{eq:nucl_int1}
\end{equation}
where $\mathcal{P}_{\pm} = \frac{1}{2}\left(\mathds{1} \pm \gamma_4 \right)$ is the parity projector. 
The matrices $\Gamma_1^{(i)}$ and $\Gamma_2^{(i)}$ are given in Table \ref{tab:BI}. 
In our case the diquark $\{d_b^T \Gamma_2^{(i)}u_c  -
u_b^T \Gamma_2^{(i)}d_c\}$ has spin $0$ and isospin $I=0$.

\begin{table}[tb]
\caption{List of Dirac structures for the $N$ baryon fields
with scalar or pseudoscalar diquarks, where $I$ is the isospin, $J^P$ indicates spin and parity. The $s_2^{(i)}$ come from the relation $\gamma_4 \Gamma_2^{(i)\;\dagger}\gamma_4 = s_2^{(i)}\Gamma_2^{(i)}$. 
}
\begin{ruledtabular}
\begin{tabular}{c|cccc}
  		$I,J^P$ & $\Gamma_1^{(i)}$ & $\Gamma_2^{(i)}$ & $s_2^{(i)}$ & $i$\\
  		\colrule 
  		& $\mathds{1}$ & $C\gamma_5$ & $+1$ & 1 \\
  		$N^{(i)}\left( \frac{1}{2},\frac{1}{2}^{\pm} \right) $ & $\gamma_5$ & $C$ & $-1$ & 2 \\
  		& $\I\mathds{1}$ & $C\gamma_5\gamma_4$ & $+1$ & 3\\
 	    & $\I\gamma_5$ & $C\gamma_4$ & $+1$ & 4\\
  	\end{tabular}\label{tab:BI}
  	\end{ruledtabular}
\end{table}

It is known that only two local nucleon operators are linearly
independent if one takes into account requirements of Lorentz- and
Fierz-invariance \cite{D}. However, the chiral-spin symmetry is not
a symmetry of the Dirac equation and the chiral-spin transformations mix 
different irreducible representations of  the Lorentz group. Consequently
if one discusses properties of  operators under the chiral-spin
transformations one needs a complete set of such operators with respect
to $SU(2)_{CS}$. Since a single-quark field transforms under a two-dimensional
irreducible representation (\ref{V-def})-(\ref{eq:su2cs_}) of $SU(2)_{CS}, k=4$  a 
complete set of three-quark nucleon interpolators with respect to $SU(2)_{CS}$
should contain eight  independent operators of positive and negative
parity because 
$\bm{2}\otimes\bm{2}\otimes\bm{2}= \bm{2}_1 \oplus \bm{2}_2 \oplus \bm{4}$.
Such operators with $J=0$ diquark are listed in Table \ref{tab:BI}.

\begin{figure}[htb]
\includegraphics[scale=0.65]{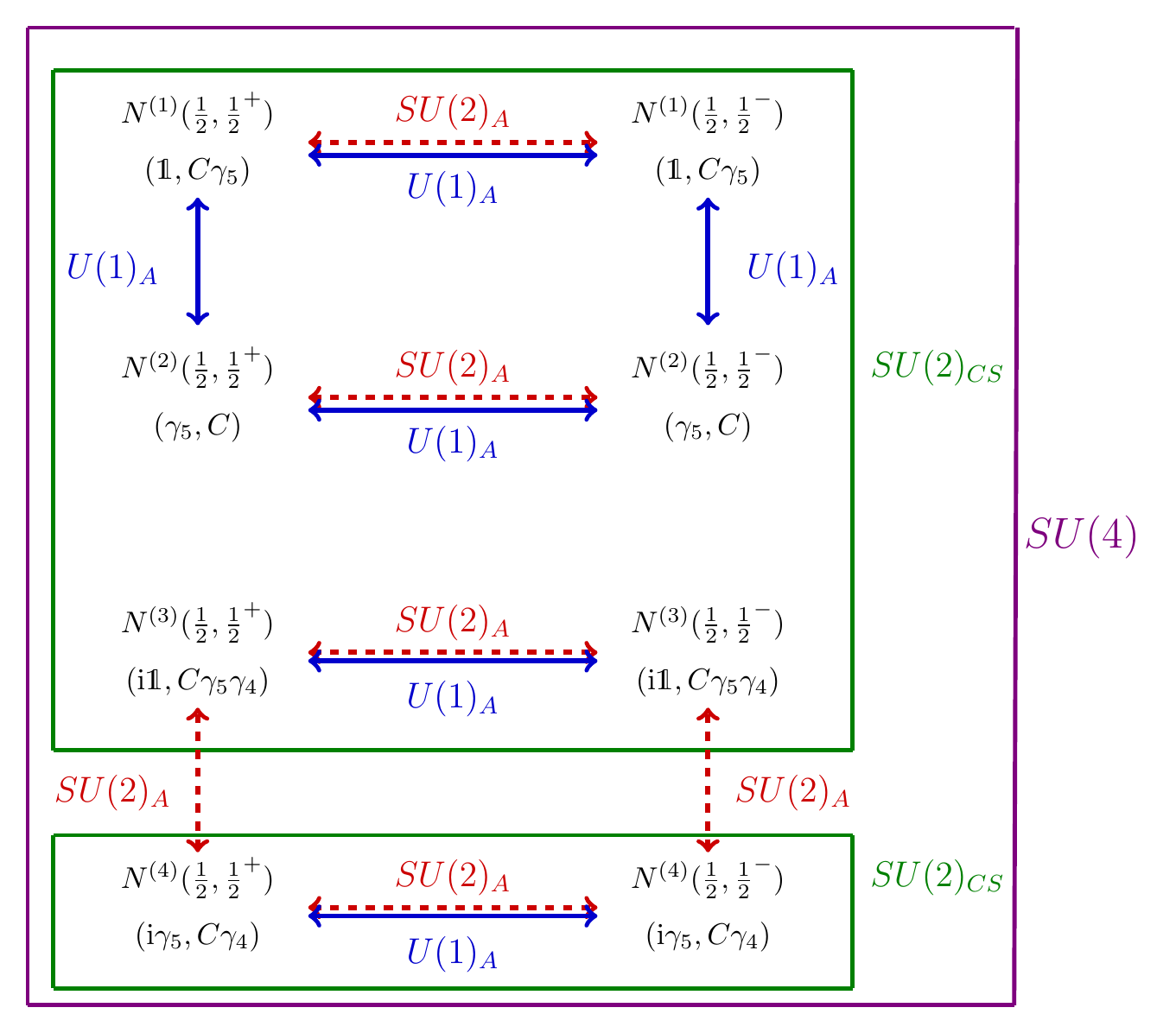}
\caption{ The nucleons linked by dashed red arrows are connected by $SU(2)_A$, 
by blue arrows are connected by $U(1)_A$. 
The nucleons inside the green boxes are all connected via $SU(2)_{CS}$ and inside the violet box are connected via $SU(4)$. 
}
\label{fig:nuclink}
\end{figure}

Applying the $U(1)_A$ transformation on the given operator of Table \ref{tab:BI}, one obtains  a linear combination of some
operators that are connected by blue arrows in Fig. \ref{fig:nuclink}.
Consequently the operators connected by blue arrows form reducible representations of $U(1)_A$. The irreducible representations of 
$U(1)_A$  are one-dimensional and can be obtained as certain linear
combinations of operators connected by blue arrows.

 The axial part of $SU(2)_L \times SU(2)_R$ (abbreviated
as $SU(2)_A$) transforms the given operator into a linear
superposition of operators connected by dashed red lines on 
Fig. \ref{fig:nuclink}. For example, both the operators of positive and negative
parity
$N^{(1)}\left( \frac{1}{2},\frac{1}{2}^{\pm} \right) $
form a four-dimensional irreducible representation $(0,1/2)+ (1/2,0)$ of the 
parity-chiral group. The same is true for the operators 
$N^{(2)}\left( \frac{1}{2},\frac{1}{2}^{\pm} \right) $.

For the operators $N^{(3)}\left( \frac{1}{2},\frac{1}{2}^{\pm} \right) $
 as well as $N^{(4)}\left( \frac{1}{2},\frac{1}{2}^{\pm} \right) $
 the situation is a bit more complicated. Applying the $SU(2)_L \times SU(2)_R$
 transformation on each of these operators one obtains   linear
 combinations of these operators as well as 
 of $\Delta$-operators (isospin $I=3/2$) of the same spin. This is because
 certain linear combinations of $N^{(3)}\left( \frac{1}{2},\frac{1}{2}^{\pm} \right) $
 and $N^{(4)}\left( \frac{1}{2},\frac{1}{2}^{\pm} \right) $ form along with
 their $\Delta$-partners the
 irreducible representations $(1,1/2)+ (1/2,1)$. 

The $SU(2)_{CS}, k=4$ transformations connect all operators inside the green boxes
of Fig. \ref{fig:nuclink}.
Finally the $SU(4)$ transformations connect all eight operators of Fig. \ref{fig:nuclink}
along with the respective $\Delta$-partners.

Below we present a set of nucleon operators that
transform under irreducible representations of $SU(2)_{CS}, k=4$ \cite{CG}.
These operators are linear combinations of the operators from
the Table \ref{tab:BI}:

\begin{equation}
\begin{split}
&B_{2_1}(-1/2) =  \frac{1}{4\sqrt{2}} \gamma_-\left[-(N^{(1)}_{+}-N^{(1)}_{-}) +(N^{(2)}_{+}-N^{(2)}_{-})- \I (N^{(3)}_{+}+N^{(3)}_{-}) +\I (N^{(4)}_{+}+N^{(4)}_{-})\right]\\
&B_{2_1}(1/2) =\frac{1}{4\sqrt{2}} \gamma_-\left[(N^{(1)}_{+}+N^{(1)}_{-}) -(N^{(2)}_{+}+N^{(2)}_{-})+ \I (N^{(3)}_{+}-N^{(3)}_{-}) -\I (N^{(4)}_{+}-N^{(4)}_{-})\right]\\
&B_{2_2}(-1/2) =\frac{1}{8}\sqrt{\frac{2}{3}}\gamma_-\left[-(N^{(1)}_{+}-N^{(1)}_{-}) + (N^{(2)}_{+}-N^{(2)}_{-}) - \I (N^{(3)}_{+}+N^{(3)}_{-}) -3\I (N^{(4)}_{+}+N^{(4)}_{-})\right]\\
&B_{2_2}(1/2) =\frac{1}{8}\sqrt{\frac{2}{3}}\gamma_-\left[ (N^{(1)}_{+}+N^{(1)}_{-}) - (N^{(2)}_{+}+N^{(2)}_{-}) + \I (N^{(3)}_{+}-N^{(3)}_{-}) +3\I (N^{(4)}_{+}-N^{(4)}_{-})\right]\\
&B_4 (-3/2) = \frac{1}{4}\gamma_-\left[(N^{(1)}_{+}+N^{(1)}_{-}) + (N^{(2)}_{+}+N^{(2)}_{-})\right]\\
&B_4 (-1/2) = \frac{1}{4}\sqrt{\frac{1}{3}}\gamma_- \left[ (N^{(1)}_{+}-N^{(1)}_{-}) - (N^{(2)}_{+}-N^{(2)}_{-}) -2\I (N^{(3)}_{+}+N^{(3)}_{-})\right]\\
&B_4 (1/2) = \frac{1}{4}\sqrt{\frac{1}{3}}\gamma_- \left[ (N^{(1)}_{+}+N^{(1)}_{-}) - (N^{(2)}_{+}+N^{(2)}_{-}) -2\I (N^{(3)}_{+}-N^{(3)}_{-})\right]\\
&B_4 (3/2) = \frac{1}{4}\gamma_-\left[(N^{(1)}_{+}-N^{(1)}_{-}) + (N^{(2)}_{+}-N^{(2)}_{-})\right]\\
\end{split}
\end{equation}

\noindent
Explicitly these operators are:

\begin{equation}
\begin{array}{ll}
B_{2_1}(-1/2) &= \epsilon_{abc}\sqrt{\frac{1}{2}}\gamma_- \left[ \gamma_4 u_a \left\lbrace d_b^{T} C \gamma_- u_c \right\rbrace+u_a \left\lbrace d_b^T  C\gamma_4 \gamma_- u_c \right\rbrace \right],\vspace{3pt}\\
B_{2_1}(1/2) &=\epsilon_{abc}\sqrt{\frac{1}{2}}\gamma_- \left[ 
u_a \lbrace d_b^T C \gamma_+ u_c \rbrace +\gamma_4 u_a \lbrace d_b^T C \gamma_4 \gamma_+ u_c \rbrace \right],\vspace{3pt}\\
B_{2_2}(-1/2) &= \epsilon_{abc}\sqrt{\frac{1}{6}}\gamma_-  \left[ 
-2 u_a \lbrace d_b^T  C\gamma_4\gamma_+ u_c\rbrace - u_a \lbrace d_b^T C\gamma_4 \gamma_-  u_c \rbrace 
+\gamma_4 u_a \lbrace d_b^T C \gamma_-  u_c \rbrace \right],\vspace{3pt}\\
B_{2_2}(1/2) &= \epsilon_{abc}\sqrt{\frac{1}{6}}\gamma_-  \left[ 
-2\gamma_4 u_a \lbrace d_b^T C \gamma_4\gamma_-  u_c\rbrace -\gamma_4 u_a \lbrace d_b^T C \gamma_4 \gamma_+  u_c \rbrace   
+ u_a \lbrace d_b^T C \gamma_+  u_c \rbrace  \right],\vspace{3pt}\\
B_{4}(-3/2) &= -\epsilon_{abc}\gamma_-   u_a \lbrace d_b^{T} C \gamma_-  u_c\rbrace, \vspace{3pt}\\
B_{4} (-1/2) &= \epsilon_{abc}\sqrt{\frac{1}{3}}\gamma_-  \left[ -u_a \lbrace d_b^T C\gamma_4 \gamma_+ u_c \rbrace  + u_a \lbrace d_b^T C\gamma_4 \gamma_- u_c \rbrace
-\gamma_4 u_a \lbrace d_b^T C \gamma_- u_c \rbrace\right],\vspace{3pt}\\
B_{4} (1/2) &= \epsilon_{abc}\sqrt{\frac{1}{3}}\gamma_-  \left[ \gamma_4 u_a \lbrace d_b^T  C\gamma_4 \gamma_- u_c \rbrace  - \gamma_4 u_a \lbrace d_b^T C\gamma_4 \gamma_+ u_c \rbrace
+ u_a \lbrace d_b^T C \gamma_+ u_c \rbrace\right],\vspace{3pt}\\
B_{4}(3/2) &= \epsilon_{abc}\gamma_-  \gamma_4 u_a \lbrace d_b^T C \gamma_+  u_c \rbrace,\\
\end{array}
\label{eq:rep_nucl}
\end{equation}

\begin{figure}[tb]
\includegraphics[scale=0.8]{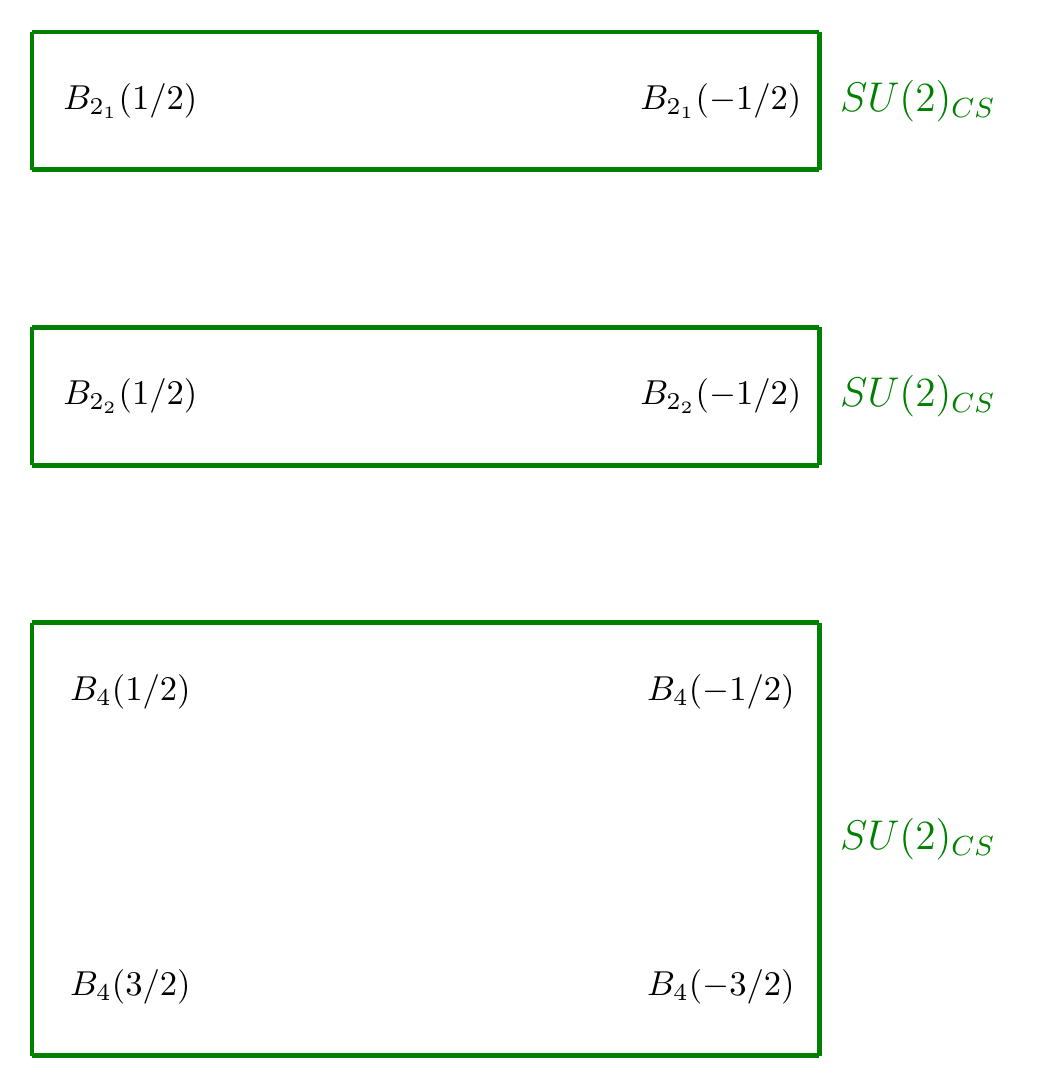}
\caption{Nucleons $B_{r} (\chi_z)$ in the irreducible representations of $SU(2)_{CS}$. Operators inside the green boxes form the basis of
the corresponding irreducible representation and are  connected via $SU(2)_{CS}$ transformations.
}
\label{fig:barlink}
\end{figure}

\noindent
Here $\gamma_\pm=\frac{1}{2}({\mathds{1}\pm\gamma_5})$ and $B_r (\chi_z)$ is the nucleon interpolator in the irreducible
representation  of dimension $r = 2\chi + 1$ of $SU(2)_{CS}$ and with chiral-spin index $\chi_z$ ($z$-projection of the chiral-spin $\chi$). 
In  (\ref{eq:rep_nucl}) the curly brackets $\{...\}$ mean antisymmetrization
between $d_b$ and $u_c$ quarks like in  (\ref{eq:nucl_int1}).
Upon
the chiral-spin transformation (\ref{V-def})-(\ref{eq:su2cs_}) with $k=4$
 only those nucleon operators are connected that belong to 
 the same irreducible representation, as illustrated in 
 Fig. \ref{fig:barlink}.

\section{Spectral decomposition {\label{sec:Baryonpropagators}}} 

In this section we analyse the Euclidean nucleon 
propagators along $t$-direction upon truncation of the low-lying modes of the
Dirac operator. 
We follow the procedure that was developed in Ref. \cite{Lang} for a similar study of meson propagators. 
This approach is based on the spectral decomposition of the quark propagator in terms of the eigenmodes of the Dirac operator. The eigenmodes contain complete information about interaction of a quark with a gluonic field. 

 We work in Euclidean space-time and  consider   a hermitian massless Dirac operator $D_0 \equiv \I\gamma_{\mu}D_{\mu}$. 
The eigenfunctions and eigenvalues of the Dirac operator are defined by the relation
\begin{equation}
D_0\psi^{(n)} = \eta_n \psi^{(n)}. 
\label{eq:dirac1}
\end{equation}
Because of $\left\lbrace  \gamma_5, D_0 \right\rbrace =0$,  the eigenvalues come in pairs with opposite signs $(\eta_n,-\eta_n)$ since 
\begin{equation}
D_0 \gamma_5 \psi^{(n)} = -\eta_n  \gamma_5\psi^{(n)}.
\label{eq:dirac2}
\end{equation} 
In the following we will use the notation: $\eta_{-n} \equiv -\eta_{n}$. 
Here and in the rest of this work we assume that the Dirac operator $D_0$ does not have exact zero modes in its spectrum, which is equivalent to selecting
gauge configurations with zero global topological charge.  The contribution of exact zero modes to observables
vanishes in the thermodynamic limit.  
Therefore in Eqs. (\ref{eq:dirac1}) and (\ref{eq:dirac2}), $\eta_n \neq 0 $, for all $\psi^{(n)} $.

The full Dirac operator for a quark field with mass $m$ can be decomposed as 
\begin{equation}
\begin{split}
&D= D_0 + \I m
= \sum_n (\eta_n + \I m) \psi^{(n)}\psi^{(n)\;\dagger} \\
& = \sum_{n>0} \left[ (\eta_n +\I m)\psi^{(n)}\psi^{(n)\;\dagger} + 
(-\eta_n + \I m)\gamma_5 \psi^{(n)}\psi^{(n)\;\dagger}\gamma_5\right], 
\end{split}
\label{eq:dirac3}
\end{equation}
where we used (\ref{eq:dirac1}) and (\ref{eq:dirac2}).

Now we consider baryon propagators and their decomposition  using 
 (\ref{eq:dirac3}) for a theory with two mass degenerate quark flavours.
A general baryon interpolator, see Eq. (\ref{eq:rep_nucl}), can be written as
\begin{equation}
 O(x) = \sum_i c_i O^{(i)}(x),
 \label{eq:baryon_int2}
\end{equation}
for some choice of the coefficients $c_i \in \mathds{C}$, in which
\begin{equation}
 O^{(i)}(x) = \epsilon_{abc}\hat{\Gamma}_1^{(i)}u_a \{d_b^T \Gamma_2^{(i)} u_c - u_b^T \Gamma_2^{(i)} d_c\},
\end{equation}
where  
$\hat{\Gamma}_1^{(i)}$ is given by a linear combination of products of Dirac matrices, 
$\Gamma_2^{(i)}$ is a generic product of gamma matrices and it satisfies the relation: $\gamma_4 \Gamma_2^{(i)\;\dagger}\gamma_4 = s_2^{(i)}\Gamma_2^{(i)}$, where $s_2^{(i)}= \pm 1$. 

The propagator  associated with the operators  $O^{(i)}(x)$ and $O^{(j)}(y)$, after the application of the Wick contractions is given by 
\begin{equation}
\begin{split}
&C^{(i,\;j)} (x,y) = \langle O^{(i)}(x) \bar{O}^{(j)}(y)\rangle_A \\
&=s_2^{(i)} \epsilon_{abc}\epsilon_{a'b'c'}(\tilde{\Gamma}_1^{(i)})_{\xi\alpha}(\Gamma_2^{(i)})_{\beta\gamma}(\Gamma_2^{(j)})_{\gamma'\beta'}(\gamma_4 \tilde{\Gamma}_1^{(j)\;\dagger})_{\alpha'\xi}\\
&\left[D^{-1}_{u_{xa\alpha|ya'\alpha'}}D^{-1}_{d_{xb\beta|yb'\beta'}}D^{-1}_{u_{xc\gamma|yc'\gamma'}}\right.\\ 
& - \left.D^{-1}_{u_{xa\alpha|yc'\gamma'}}D^{-1}_{d_{xb\beta|yb'\beta'}}D^{-1}_{u_{xc\gamma|ya'\alpha'}}\right]
\end{split}
\label{eq:propag_gen}
\end{equation}

Furthermore we have called, e.g., $D^{-1}_{u_{xa\alpha|ya'\alpha'}} = \langle u_{xa\alpha}\; \bar{u}_{ya'\alpha'}\rangle_A$ the quark propagator of the $up$ quark between the space-time points  $x$ and $y$, with colour indices $a$ and $a'$, and Dirac indices $\alpha$ and $\alpha'$. 
In the case of two degenerate quark masses, then $D^{-1} \equiv D^{-1}_u = D^{-1}_d$.

In absence of zero modes in the Dirac spectrum, the quark propagator $D^{-1}$  can be expanded (see Ref. \cite{Lang} and Eq. (\ref{eq:dirac3})) as 
\begin{equation}
\begin{split}
&D^{-1}_{xa\delta|ya'\alpha'} = \sum_{n>0} f_n\; \psi^{(n)}_{xa\alpha} \psi^{(n)\;\dagger}_{ya'\alpha'} + f_{-n}\;(\gamma_5)_{\alpha\xi}\psi^{(n)}_{xa\xi} \psi^{(n)\;\dagger}_{ya'\xi'} (\gamma_5)_{\xi'\alpha'}
\end{split}
\label{eq:prop1}
\end{equation}
where
\begin{equation}
\begin{split}
  &f_n = \frac{1}{\eta_n +\I m} = h_n -  \I g_n \\
  &f_{-n} = \frac{1}{\eta_{-n} +\I m} = -h_n - \I g_n\\
\end{split}  
\label{eq:f}
\end{equation}
with
\begin{equation}
\begin{split}
&h_n\equiv h(m,\eta_n) = \frac{\eta_n}{m^2 + \eta_n^2}\\
&g_n\equiv g(m,\eta_n) = \frac{m}{m^2 + \eta_n^2},\qquad n>0.\\
 \end{split}
\end{equation}
Substituting the Eq. (\ref{eq:prop1}) in the full propagator: 
\begin{equation}
C(x,y) = \langle O(x) \bar{O}(y)\rangle_A  =\sum_{i,j} c_i c_j^{*} \; C^{(i,\;j)} (x,y) ,
\label{prop}
\end{equation}
we can express it in terms of $h(m,\eta)$ and $g(m,\eta)$, 
\begin{equation}
\begin{split}
C(x,y) &= \sum_{n>0, k>0, l>0}  \left( g_n g_k g_l S^{ggg}(x,y)+
 g_n g_k h_l S^{ggh}(x,y) \right.  \\
 &\left. +g_n h_k h_l S^{ghh}(x,y) +
 h_n h_k h_l S^{hhh}(x,y)\right), \;.
 \end{split}
 \label{eq:prop_decomp}
\end{equation}
The functions $S^{ggg}(x,y)$,$S^{ggh}(x,y)$, $S^{ghh}(x,y)$  and $S^{hhh}(x,y)$ contain the information about the eigenfunctions of the Dirac operator and the structure of the baryon field under consideration.

\begin{figure}[tb]
\includegraphics[scale=1.0]{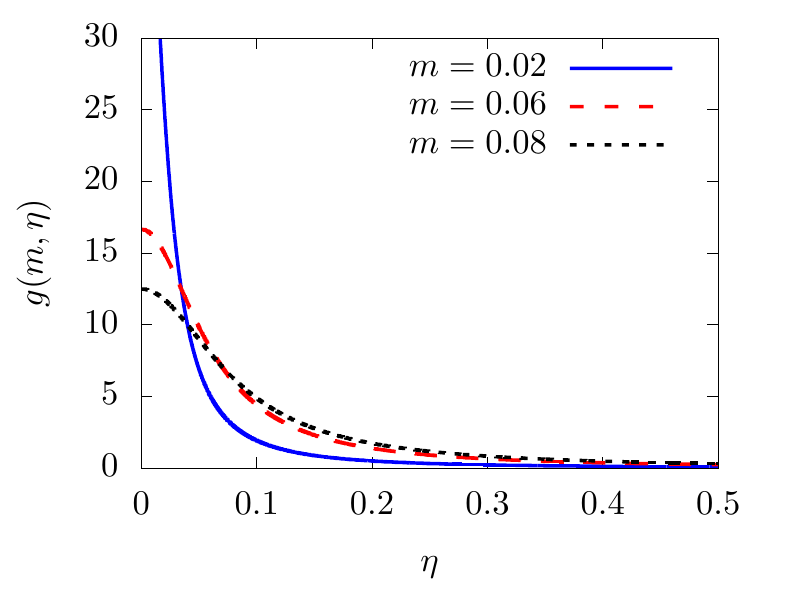}
\includegraphics[scale=1.0]{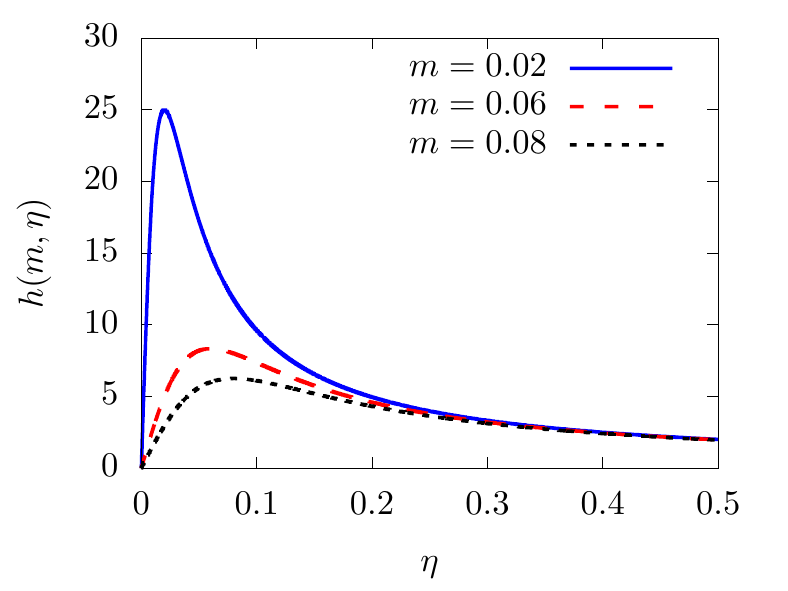}
\caption{$g(m,\eta)$ and $h(m,\eta)$ functions from Ref. \cite{Lang} for $m=0.02$ (full), $0.06$ (dashed) and $0.08$ (dotted).
}
\label{fig:gh}
\end{figure}

Therefore the correlator $C(x,y)$ has terms proportional to the $g(m,\eta)$ function, like $g_n g_k g_l S^{ggg}(x,y)$, $g_n g_k h_l S^{ggh}(x,y)$ and $g_n h_k h_l S^{ghh}(x,y)$, that we call \textit{$g$-terms}, and  terms proportional only to the $h(m,\eta)$ function, that we call \textit{$h$-terms}.
A sketch of these two functions for different mass values is shown in Fig. \ref{fig:gh}.

In the chiral limit $m\rightarrow 0 $ the function  $g(m,\eta)$ approaches the delta-function
  $\frac{\pi}{2}\delta(\eta)$.
  Hence a  gap around zero in the spectrum of the Dirac operator
  will induce vanishing of the terms in Eq. (\ref{eq:prop_decomp})
  that contain at least one factor of $g$. In other words, all
  $g$-{\it terms} in Eq. (\ref{eq:prop_decomp}) vanish in the chiral
  limit upon truncation of the near-zero modes of the Dirac operator.

The 
$h(m,\eta)$ function is peaked at $\eta = m$ and  
falls slower compared to the  $g(m,\eta)$ function at high eigenvalues 
$\eta$. Consequently while the $h(m,\eta)$ function still suppresses higher
eigenvalues $\eta$, making a small hole in the Dirac eigenspectrum will
not necessarily lead to the vanishing of the $h$-term in Eq. (\ref{eq:prop_decomp})
in the chiral limit
unless some additional  suppressing dynamical factors are contained in $S^{hhh}(x,y)$. 
 
In the following we  call nucleon operators
\textit{$g$-equivalent}
if  the difference of their propagators contains only \textit{$g$-terms}. 

\section{Spectral decomposition of nucleon propagators{\label{subsec:Nucleonpropagator}}}
\subsection{Correlators of $N^{(i)}$ operators{\label{subsec:Nucleonpropagator_gh}}}

Now we apply results of the previous section to correlators of nucleon operators
from  Table \ref{tab:BI}.
The details of the expansion in \textit{$g$-terms} and \textit{$h$-term} of the nucleon propagators are given in Appendix \ref{app:A}.\\

In Fig. \ref{fig:nucgh} we show how the difference of two correlators (\ref{eq:prop_decomp})
calculated with any two operators from  Table \ref{tab:BI} is expressed
via the $ggg$, $ggh$, $ghh$  and $hhh$ terms.
\begin{figure}[tb]
\includegraphics[scale=0.8]{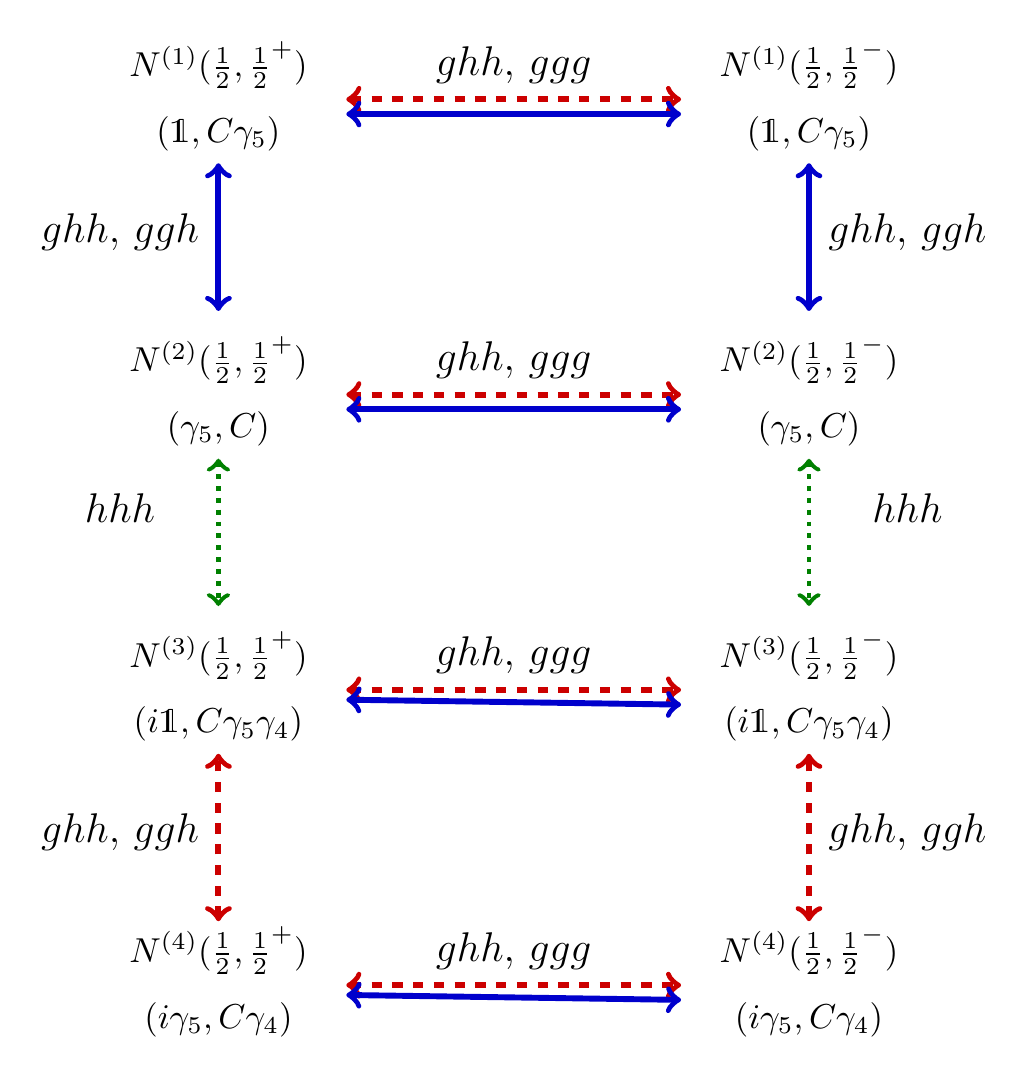}
\caption{$g$ and $h$ connections among nucleons of Table \ref{tab:BI}. 
Below each nucleon we have indicated its $\Gamma$ structure, i.e. $(\Gamma_1^{(i)},\Gamma_2^{(i)})$.
}
\label{fig:nucgh}
\end{figure}
We see from Fig. \ref{fig:nucgh} that all nucleons connected by $U(1)_A$ and/or $SU(2)_A$
transformations, see Fig. \ref{fig:nuclink}, are \textit{$g$-equivalent}, see for details Appendices \ref{app:B} and \ref{app:C}. Consequently a gap in the
low-lying spectrum of the Dirac operator results (in the chiral limit) in degeneracy
 of all correlators obtained with operators connected by dashed red and/or blue
 arrows in Fig. \ref{fig:nuclink}.  We conclude that a gap in the Dirac spectrum
 implies necessarily restoration of both $U(1)_A$ and $SU(2)_L \times SU(2)_R$
 symmetries in nucleons. It is similar to the results for  meson
 correlators obtained in \cite{Lang}. Such degeneracies of the nucleon correlators have
 been observed on the lattice in Ref. \cite{Denissenya:2015woa}.
 
 Let us summarise. Restoration of $U(1)_A$ and $SU(2)_L \times SU(2)_R$ symmetries in nucleon correlators (\ref{prop}) is necessarily provided by a
 gap in spectrum of the Dirac operator,
  i.e. all $U(1)_A$ and $SU(2)_L \times SU(2)_R$ breaking dynamics
 is contained only in the near-zero modes.

 However, the observations of Ref. \cite{Denissenya:2015woa} went essentially
 further than simply $U(1)_A$ and $SU(2)_L \times SU(2)_R$ restoration. It
 was noticed that a larger symmetries $SU(2)_{CS}$ and $SU(4)$ emerge in
 baryon masses upon low-modes truncation. 
 
 From the analytical side we can now conclude the following.
 Comparing  Fig. \ref{fig:nucgh} with  Fig.\ref{fig:nuclink}
 we observe that, given a gap in the Dirac spectrum,  emergence of $SU(2)_{CS}$ and $SU(4)$ 
 requires in addition that the $h$-{\it term} in the difference of two correlators connected
 by the $SU(2)_{CS}$ transformation (and not connected by the
 chiral transformations) should be at
 least strongly suppressed for  higher-lying eigenmodes of the Dirac
 operator. While some suppression is indeed provided by the $hhh$ factor,
 see Fig. \ref{fig:gh}, this suppression is not as strong as in $g$-{\it terms}. 
 In other words, a gap in the Dirac spectrum does not 
 automatically imply emergence
 of the $SU(2)_{CS}$ and $SU(4)$  symmetries in correlators 
 (\ref{eq:prop_decomp}).
 
 This result is not unexpected. In contrast to the chiral symmetries the $SU(2)_{CS}$ and $SU(4)$ symmetries
 are not covariant. They are symmetries of the quark-electric interaction
 in the given reference frame, while the quark kinetic term as well as
 the quark-magnetic interaction break them. They have been observed
 as symmetries of hadron masses upon low-mode truncation, i.e. symmetries of the rest frame correlation functions. The correlators (\ref{prop}) mix different
 reference frames in Minkowski space and only covariant symmetries, such as chiral symmetries, should persist in these correlators. Consequently to address the question
 of symmetries of hadron masses we need now to analyze the rest frame correlators (\ref{eq:momentumprojection}).
 This means we need to study the correlators 
 \begin{equation}
 C^{(i)}_{\pm}(t) = \sum_{x,y,z} \langle N^{(i)}_{\pm}(x,y,z,t)\bar{N}^{(i)}_{\pm}(\bm{0},0)\rangle,
\label{rf}
 \end{equation}
 where the sum $\sum_{x,y,z}$ is over the all space. 
 
 However summation over all spatial points $x,y,z$ does not
 convert an $h$-connection between the $N^{(2)}$  and $N^{(3)}$ 
 operators in Fig. \ref{fig:nucgh}
 into a  $g$-connection.  
 We do not get further $g$-equivalence as compared the ones indicated in Fig. \ref{fig:nucgh}. The presence of a gap in the Dirac
 spectrum does not automatically imply emergence of
 $SU(2)_{CS}$ and $SU(4)$.
 
 In full QCD studies with the explicite removal of the low lying modes
in the propagators the $SU(2)_{CS}$ and $SU(4)$ symmetries were observed in the hadron spectrum  \cite{Denissenya:2015woa} .
This implies that a cancellation of $hhh$-terms occurs due to
some additional $SU(2)_{CS}$ and $SU(4)$-symmetric
microscopic dynamics, i.e., QCD dynamics
beyond the chiral symmetry breaking dynamics dominated
by the low modes. Such dynamics, as it follows from the
symmetry classification of the QCD Lagrangian in Sec. II,
should be related with the confining quark - chromo-electric
interaction.

 Let us summarise. Restoration of $U(1)_A$ and $SU(2)_L \times SU(2)_R$ symmetries 
 in nucleon correlators (\ref{prop}) is necessarily provided by a
 gap in spectrum of the Dirac operator,
  i.e. all $U(1)_A$ and $SU(2)_L \times SU(2)_R$ breaking dynamics
 is contained only in the near-zero modes.
 The $SU(2)_{CS}$ and $SU(4)$ symmetries in the rest-frame correlators
 (\ref{rf}) do not automatically emerge. Their emergence 
 requires some additional microscopical dynamical input that would guarantee that contributions of the high-lying
 modes is  $SU(2)_{CS}$ and $SU(4)$ symmetric.

 \subsection{$B_r (\chi_z)$ baryon propagators}
 
 In Fig. \ref{fig:barlink} we have reported irreducible  $SU(2)_{CS}$ representations 
 of the baryon operators defined in Eq. (\ref{eq:rep_nucl}). 
 Each of these operators is a $U(1)_A$-singlet, i.e. transforms into
 itself upon the $U(1)_A$ transformation. This is because
  by definition 
 the $B_r (\chi_z)$ interpolators are eigenstates of $\gamma_5$ in the different representations $\bm{2}_1$, $\bm{2}_2$ and $\bm{4}$ 
 of $SU(2)_{CS}$.

Regarding the $SU(2)_L \times SU(2)_R$, each operator from Fig. \ref{fig:barlink}
is a linear combination of positive and negative parity operators
(\ref{eq:nucl_int1}).
Different operators (\ref{eq:nucl_int1}) belong to different
irreducible representations of the parity-chiral group,
as was discussed above, so no
definite  representation of $SU(2)_L \times SU(2)_R$
can be ascribed to the operators (\ref{eq:rep_nucl}).

Now we apply a spectral decomposition  of Sec. IV
 to the propagators built with the baryon operators
 (\ref{eq:rep_nucl})
 \begin{equation}
 C(x,y)_{r,\chi_z} = \langle B_r (\chi_z)(x) \bar{B}_r (\chi_z)(y) \rangle.
 \end{equation}
 We find that the difference between two generic propagators $C(x,y)_{r,\chi_z}$ and $C(x,y)_{r',\chi_z'}$, always contains $hhh$-terms.
 This means that a gap in the spectrum of the Dirac operator
 does not yet automatically imply emergence of the $SU(2)_{CS}$ and $SU(4)$
 symmetries. 
 This result is not unexpected since the correlators $C(x,y)_{r,\chi_z}$
 mix different reference frames in Minkowski space-time and only covariant symmetries can persist
 in such correlators. It is
 in complete agreement with the result obtained for the nucleon propagators, see Fig. \ref{fig:nucgh}.
 
 Consequently we analyze now baryon correlators in the rest frame, i.e. we 
  consider the correlators
 \begin{equation}
 C(t)_{r,\chi_z} = \sum_{x,y,z}\langle B_r (\chi_z)(x,y,z,t) \bar{B}_r (\chi_z)(\bm{0},0) \rangle,
 \label{eq:ct_b}
 \end{equation}
where the sum $\sum_{x,y,z}$ is over the all space.

 Under parity transformations the quark fields transform as 
\begin{equation}
q_{xa\alpha}\xrightarrow{P}q_{xa\alpha}^P= (\gamma_4)_{\alpha\beta}q_{\mathcal{P}xa\beta}
\label{eq:parity_quark}
\end{equation}
where $\mathcal{P}x \equiv \mathcal{P}_{\mu\nu}x_{\nu}$ and $\mathcal{P}_{\mu\nu} = diag(-1,-1,-1,1)$ is the parity operator, hence if $x_{\mu}=(x,y,z,t)$, then $(\mathcal{P}x)_{\mu} = (-x,-y,-z,t)$. $q$ is a generic quark field. 
Applying the parity transformations  (\ref{eq:parity_quark}) to the baryon operators in (\ref{eq:rep_nucl}) we get the following relation for generic representation $r$ and chiralspin projection $\chi_z$
\begin{equation}
B_r(\chi_z)(\mathcal{P}x)=  \eta B_r^P(-\chi_z)(x),
\label{eq:parity_b}
\end{equation}
where we indicate $B_r(\chi_z)(\mathcal{P}x)=B_r(\chi_z)(-x,-y,-z,t)$ and 
$B_r^P (\chi_z)$ is the baryon operator $B_r (\chi_z)$ in (\ref{eq:rep_nucl}), and we have substitued $u\rightarrow u^P$ and $d\rightarrow d^P$, see (\ref{eq:parity_quark}).
In Eq. (\ref{eq:parity_b}) $\eta = \pm 1$, depending on $r$ and $\chi_z$, 
and we used that $\gamma_4 C = -C\gamma_4$ and that $\gamma_4 \gamma_{\pm}\gamma_4 = \gamma_{\mp}$.
Plugging Eq. (\ref{eq:parity_b}) in (\ref{eq:ct_b}) we get 
\begin{equation}
\begin{split}
 C(t)_{r,\chi_z} &= \sum_{x,y,z}\langle B_r (\chi_z)(x,y,z,t) \bar{B}_r (\chi_z)(\bm{0},0) \rangle\\
 &=\sum_{x,y,z}\langle B_r (\chi_z)(-x,-y,-z,t) \bar{B}_r (\chi_z)(\bm{0},0) \rangle\\
 &=\sum_{x,y,z}\langle B_r^P (-\chi_z)(x,y,z,t) \bar{B}_r^P(-\chi_z)(\bm{0},0) \rangle\\
 &=\sum_{x,y,z}\langle B_r(-\chi_z)(x,y,z,t) \bar{B}_r(-\chi_z)(\bm{0},0) \rangle=C(t)_{r,-\chi_z},
 \end{split}
 \label{eq:chi_eq}
 \end{equation}
 where   in the third line we used Eq. (\ref{eq:parity_b}).
  Since we are averaging over all possible quark fields we can remove 
  the label $P$  in the last line of Eq. (\ref{eq:chi_eq})
 (because parity is a symmetry of the QCD action and the measure in the average $\langle\cdot\rangle$ is parity-invariant). 

  Eq. (\ref{eq:chi_eq}) tells us that for a given irreducible representation $r$ of $SU(2)_{CS}$ we have $ C(t)_{r,\chi_z} -  C(t)_{r,-\chi_z}=0$, for all $\chi_z$. 
 Hence in the rest frame the correlators for the baryons within
the doublet $\bm{2}_1$ and $\bm{2}_2$ representations are equal.
This is a general statement, irrespective whether there is or there is not
a gap in the spectrum of the Dirac operator. This fact does not
mean, however, that the $SU(2)_{CS}$ symmetry is manifest in the rest-frame
correlators, because in the representation $\bm{4}$ the correlators
with $\chi_z = \pm 1/2$ are not equal to the correlators with
$\chi_z = \pm 3/2$.
  
A presence of a gap in the Dirac spectrum does not automatically make
 the correlators
with $\chi_z = \pm 1/2$ and  with
$\chi_z = \pm 3/2$ \textit{$g$-equivalent}. 
The emergence 
of $SU (2)_{CS}$ requires some additional suppression of 
matrix elements with higher-lying modes as was discussed
in the previous subsection.

\section{Conclusions}

In this paper we have analysed analytically, by
expansion of the propagators into eigenmodes
of the Dirac operator, which symmetries emerge in  baryon 
correlators (masses) if there is a gap around zero
in the spectrum of the Dirac operator. We have found that such 
a gap results necessarily in emergence of chiral $U(1)_A$ and
$SU(2)_L \times SU(2)_R$ symmetries in baryons. 

Some specific dynamics in QCD leads to the accumulation of the
near-zero modes, i.e. to the breaking of chiral symmetries.
Given the $\gamma^5$-anticommutativety of the Euclidean Dirac
operator we prove here that a gap in the Dirac eigenmode spectrum
implies necessarily restoration of both $U(1)_A$ and
$SU(2)_L \times SU(2)_R$ symmetries. The root of this statement
is precisely the same as of Banks-Casher relation.
We do not need to know
which dynamics and why it leads to the accumulation of the 
near-zero modes.

Emergence
of larger $SU (2)_{CS}$ and $SU(4)$ symmetries, that was observed on
the lattice upon truncation of the near-zero modes of the Dirac
operator and also at high temperatures without any truncation,
requires that
the electric interaction should be the most important
for higher-lying modes. This is burried in the eigenfunctions
of the Dirac operator and cannot be specified within the
present approach which does not use any dynamical input.

\begin{acknowledgments}
Support from the Austrian Science Fund (FWF) through the grants
DK W1203-N16 and P26627-N27 is acknowledged.
\end{acknowledgments}

\newpage

\appendix

\section{Nucleon propagator expansion}\label{app:A}
Using Eq. (\ref{eq:propag_gen}) and the expansion of the quark propagator in (\ref{eq:prop1}), we can get the expansion of the propagator for the nucleon interpolators given in Eq. (\ref{eq:nucl_int1}) and specified Table \ref{tab:BI}. It is given by 
\begin{equation}
\begin{split}
&C(N_{\pm}^{(i)}) = s_2^{(i)}\epsilon_{abc} \epsilon_{a'b'c'}(\gamma_4\Gamma_1^{(i)\;\dagger})_{\alpha'\omega}  (\mathcal{P}_{\pm})_{\omega\epsilon}(\Gamma_1^{(i)})_{\epsilon\alpha}\\
&(\Gamma_2^{(i)})_{\beta\gamma}(\Gamma_2^{(i)})_{\gamma'\beta'}\left[D^{-1}_{xa\alpha|ya'\alpha'}D^{-1}_{xb\beta|yb'\beta'}D^{-1}_{xc\gamma|yc'\gamma'}\right.
 - \left.D^{-1}_{xa\alpha|yc'\gamma'}D^{-1}_{xb\beta|yb'\beta'}D^{-1}_{xc\gamma|ya'\alpha'}\right]\;.
\end{split}
\label{eq:corr_expa0}
\end{equation}

The last line of Eq. (\ref{eq:corr_expa0}) can be written as the following sum,
\begin{equation}
\begin{split}
D^{-1}_{xa\alpha|ya'\alpha'}&D^{-1}_{xb\beta|yb'\beta'}D^{-1}_{xc\gamma|yc'\gamma'}-D^{-1}_{xa\alpha|yc'\gamma'}D^{-1}_{xb\beta|yb'\beta'}D^{-1}_{xc\gamma|ya'\alpha'}\\
=&\sum_{n>0, k>0, l>0} \Big[
f_n f_k f_l	\left[
\psi^{(n)}_{xa\alpha} \psi^{(n)\;\dagger}_{ya'\alpha'}
\psi^{(k)}_{xb\beta} \psi^{(k)\;\dagger}_{yb'\beta'}
\psi^{(l)}_{xc\gamma} \psi^{(l)\;\dagger}_{yc'\gamma'}\right]\\
+&f_n f_k f_{-l}\left[
\psi^{(n)}_{xa\alpha} \psi^{(n)\;\dagger}_{ya'\alpha'}
\psi^{(k)}_{xb\beta} \psi^{(k)\;\dagger}_{yb'\beta'}
(\gamma_5)_{\gamma\theta}\psi^{(l)}_{xc\theta} \psi^{(l)\;\dagger}_{yc'\theta'}(\gamma_5)_{\theta'\gamma'}\right]\\		
+&f_n f_{-k} f_l\left[
\psi^{(n)}_{xa\alpha} \psi^{(n)\;\dagger}_{ya'\alpha'}
(\gamma_5)_{\beta\omega}\psi^{(k)}_{xb\omega} \psi^{(k)\;\dagger}_{yb'\omega'}(\gamma_5)_{\omega'\beta'}
\psi^{(l)}_{xc\gamma} \psi^{(l)\;\dagger}_{yc'\gamma'}\right]\\
+&f_{-n} f_k f_l\left[
(\gamma_5)_{\alpha\xi}\psi^{(n)}_{xa\xi} \psi^{(n)\;\dagger}_{ya'\xi'}(\gamma_5)_{\xi'\alpha'}
\psi^{(k)}_{xb\beta} \psi^{(k)\;\dagger}_{yb'\beta'}
\psi^{(l)}_{xc\gamma} \psi^{(l)\;\dagger}_{yc'\gamma'}\right]\\
+&f_n f_{-k} f_{-l}\left[
\psi^{(n)}_{xa\alpha} \psi^{(n)\;\dagger}_{ya'\alpha'}
(\gamma_5)_{\beta\omega}\psi^{(k)}_{xb\omega} \psi^{(k)\;\dagger}_{yb'\omega'}(\gamma_5)_{\omega'\beta'}(\gamma_5)_{\gamma\theta}\psi^{(l)}_{xc\theta} \psi^{(l)\;\dagger}_{yc'\theta'}(\gamma_5)_{\theta'\gamma'}\right]\\
+&f_{-n} f_k f_{-l}	\left[
(\gamma_5)_{\alpha\xi}\psi^{(n)}_{xa\xi} \psi^{(n)\;\dagger}_{ya'\xi'}(\gamma_5)_{\xi'\alpha'}
\psi^{(k)}_{xb\beta} \psi^{(k)\;\dagger}_{yb'\beta'}(\gamma_5)_{\gamma\theta}\psi^{(l)}_{xc\theta} \psi^{(l)\;\dagger}_{yc'\theta'}(\gamma_5)_{\theta'\gamma'}\right]\\
+&f_{-n} f_{-k} f_l\left[
(\gamma_5)_{\alpha\xi}\psi^{(n)}_{xa\xi} \psi^{(n)\;\dagger}_{ya'\xi'}(\gamma_5)_{\xi'\alpha'}
(\gamma_5)_{\beta\omega}\psi^{(k)}_{xb\omega}\psi^{(k)\;\dagger}_{yb'\omega'}(\gamma_5)_{\omega'\beta'}
\psi^{(l)}_{xc\gamma} \psi^{(l)\;\dagger}_{yc'\gamma'}\right]\\
+&f_{-n} f_{-k} f_{-l}\left[
(\gamma_5)_{\alpha\xi}\psi^{(n)}_{xa\xi} \psi^{(n)\;\dagger}_{ya'\xi'}(\gamma_5)_{\xi'\alpha'}
(\gamma_5)_{\beta\omega}\psi^{(k)}_{xb\omega} \psi^{(k)\;\dagger}_{yb'\omega'}(\gamma_5)_{\omega'\beta'}
(\gamma_5)_{\gamma\theta}\psi^{(l)}_{xc\theta} \psi^{(l)\;\dagger}_{yc'\theta'}(\gamma_5)_{\theta'\gamma'}\right]\\
&-(\textrm{same terms as above with }\alpha'\leftrightarrow\gamma'\textrm{ and }a'\leftrightarrow c')\Big] \;.\\
\end{split}
\label{eq:corr_expa}
\end{equation}
Using (\ref{eq:f}) we can rewrite the coefficients
in front of the eigenfunction products in (\ref{eq:corr_expa}) in terms of $g_n$ and $h_n$, i.e. 
\begin{equation}
\begin{split} 
f_n f_k f_l = \I\, g_n g_k g_l - h_ng_k g_l - g_n h_k g_l -\I\, h_n h_k g _h
 -g_n g_k h_l -\I \,h_n g_k h_l - \I \,g_n h_k h_l +h_n h_k h_l,\\
\end{split}
\label{eq:fff}
\end{equation}
moreover other coefficients can be found exploiting that $f_{-n} = -f_{n}^{*}$, see Eq. (\ref{eq:f}). 
Therefore by linearity of (\ref{eq:corr_expa}), we can get the expression of $C(N_{\pm}^{(i)})$ in terms proportional to $g_n g_k g_l$, $g_n g_k h_l$ , $g_n h_k h_l$ and $h_n h_k h_l$.

\section{$C(N_{+}^{(i)}) - C(N_{-}^{(i)})$}\label{app:B}

The difference $C(N_{+}^{(i)}) - C(N_{-}^{(i)})$ can be written using (\ref{eq:corr_expa0}) as 
\begin{equation}
\begin{split}
C(N_{+}^{(i)}) -  C(N_{-}^{(i)})&=s_2^{(i)}\epsilon_{abc} \epsilon_{a'b'c'}\,
(\gamma_4\Gamma_1^{(i)\;\dagger}\gamma_4)_{\alpha'\omega} ((\mathcal{P}_{+})_{\omega\epsilon}+(\mathcal{P}_{-})_{\omega\epsilon}(\Gamma_1^{(i)})_{\epsilon\alpha}(\Gamma_2^{(i)})_{\beta\gamma}(\Gamma_2^{(i)})_{\gamma'\beta'}\\
& \quad\left[ D^{-1}_{xa\alpha|ya'\alpha'}  D^{-1}_{xb\beta|yb'\beta'} D^{-1}_{xc\alpha'|yc'\gamma'}\right.  
\left. -D^{-1}_{xa\alpha | yc'\gamma'} D^{-1}_{xb\beta|yb'\beta'} D^{-1}_{xc\alpha'|ya'\alpha'} \right]\\
&=(-1)^{i+1}s_2^{(i)}\;\epsilon_{abc} \epsilon_{a'b'c'}\delta_{\alpha'\alpha}(\Gamma_2^{(i)})_{\beta\gamma}(\Gamma_2^{(i)})_{\gamma'\beta'}\\
&  \quad\left[ D^{-1}_{xa\alpha|ya'\alpha'}  D^{-1}_{xb\beta|yb'\beta'} D^{-1}_{xc\alpha'|yc'\gamma'}\right.
 -\left. D^{-1}_{xa\alpha | yc'\gamma'} D^{-1}_{xb\beta|yb'\beta'} D^{-1}_{xc\alpha'|ya'\alpha'} \right],\\
\end{split}
\end{equation}
where we used that $\gamma_4\mathcal{P}_{\pm}=\pm\mathcal{P}_{\pm}$, $(\mathcal{P}_{+})_{\omega\epsilon}+(\mathcal{P}_{-})_{\omega\epsilon} = \delta_{\omega\epsilon}$ and that $\Gamma_1^{(i)\;\dagger}\Gamma_1^{(i)}= \mathds{1}$, for all values of $i$, see Table \ref{tab:BI}.

\goodbreak
We expand the quark propagator according to (\ref{eq:corr_expa}) and use $\psi^{(-n)}=\gamma_5\psi^{(n)}$ and $\gamma_5 \Gamma_2^{(i)} \gamma_5 = s_{5(i)} \Gamma_2^{(i)}$ with $s_{5(i)}^2 = 1$, to get 
\begin{equation}
\begin{split}
&\hspace*{-15mm}C(N_{+}^{(i)}) -  C(N_{-}^{(i)})=
(-1)^{i+1} s_2^{(i)}\epsilon_{abc} \epsilon_{a'b'c'}\delta_{\alpha'\alpha}(\Gamma_2^{(i)})_{\beta\gamma}(\Gamma_2^{(i)})_{\gamma'\beta'}\\
&\hspace*{-15mm}\sum_{n>0, k>0, l>0} 
\Big[(f_n f_k f_l + f_{-n} f_{-k} f_{-l})\\&
\left[
\psi^{(n)}_{xa\alpha} \psi^{(n)\;\dagger}_{ya'\alpha'}
\psi^{(k)}_{xb\beta} \psi^{(k)\;\dagger}_{yb'\beta'}
\psi^{(l)}_{xc\gamma} \psi^{(l)\;\dagger}_{yc'\gamma'}\right.
\left.-\psi^{(n)}_{xa\alpha} \psi^{(n)\;\dagger}_{yc'\gamma'}
\psi^{(k)}_{xb\beta} \psi^{(k)\;\dagger}_{yb'\beta'}
\psi^{(l)}_{xc\gamma} \psi^{(l)\;\dagger}_{ya'\alpha'}
\right]\\
&+(f_n f_k f_{-l} + f_{-n} f_{-k} f_l)\\&
\left[
\psi^{(n)}_{xa\alpha} \psi^{(n)\;\dagger}_{ya'\alpha'}
\psi^{(k)}_{xb\beta} \psi^{(k)\;\dagger}_{yb'\beta'}
(\gamma_5)_{\gamma\theta}\psi^{(l)}_{xc\theta} \psi^{(l)\;\dagger}_{yc'\theta'}(\gamma_5)_{\theta'\gamma'}\right.
\left.-\psi^{(n)}_{xa\alpha} \psi^{(n)\;\dagger}_{yc'\gamma'}
\psi^{(k)}_{xb\beta} \psi^{(k)\;\dagger}_{yb'\beta'}
(\gamma_5)_{\gamma\theta}\psi^{(l)}_{xc\theta} \psi^{(l)\;\dagger}_{ya'\theta'}(\gamma_5)_{\theta'\alpha'}\right]\nonumber\\
&+(f_n f_{-k} f_l + f_{-n} f_k f_{-l})\\&
\left[\psi^{(n)}_{xa\alpha} \psi^{(n)\;\dagger}_{ya'\alpha'}
(\gamma_5)_{\beta\omega}\psi^{(k)}_{xb\omega} \psi^{(k)\;\dagger}_{yb'\omega'}(\gamma_5)_{\omega'\beta'}
\psi^{(l)}_{xc\gamma} \psi^{(l)\;\dagger}_{yc'\gamma'}\right.
\left.-\psi^{(n)}_{xa\alpha} \psi^{(n)\;\dagger}_{yc'\gamma'}
(\gamma_5)_{\beta\omega}\psi^{(k)}_{xb\omega} \psi^{(k)\;\dagger}_{yb'\omega'}(\gamma_5)_{\omega'\beta'}
\psi^{(l)}_{xc\gamma} \psi^{(l)\;\dagger}_{ya'\alpha'}\right]\\
&+(f_{-n} f_k f_l + f_n f_{-k} f_{-l})\\&
\left[(\gamma_5)_{\alpha\xi}\psi^{(n)}_{xa\xi} \psi^{(n)\;\dagger}_{ya'\xi'}(\gamma_5)_{\xi'\alpha'}
\psi^{(k)}_{xb\beta} \psi^{(k)\;\dagger}_{yb'\beta'}
\psi^{(l)}_{xc\gamma} \psi^{(l)\;\dagger}_{yc'\gamma'}\right.
\left. -(\gamma_5)_{\alpha\xi}\psi^{(n)}_{xa\xi} \psi^{(n)\;\dagger}_{yc'\xi'}(\gamma_5)_{\xi'\gamma'}
\psi^{(k)}_{xb\beta} \psi^{(k)\;\dagger}_{yb'\beta'}
\psi^{(l)}_{xc\gamma} \psi^{(l)\;\dagger}_{ya'\alpha'}\right]\Big]\,.
\end{split}
\end{equation}
Using (\ref{eq:f}) the coefficients in front of the eigenfunction products can be written as 
\begin{equation}
\begin{split}
f_n f_k f_l + f_{-n} f_{-k} f_{-l} 
&= 2\,\I\, (g_n g_k g_l -h_n h_k g _h - h_n g_k h_l -g_n h_k h_l),\\
f_n f_k f_{-l} + f_{-n} f_{-k} f_l
& = 2\, \I\, (g_n g_k g_l -h_n h_k g _h + h_n g_k h_l +g_n h_k h_l),\\
f_n f_{-k} f_l + f_{-n} f_{k} f_{-l}
&= 2\, \I\, (g_n g_k g_l +h_n h_k g _h - h_n g_k h_l +g_n h_k h_l),\\
f_{-n} f_k f_l + f_{n} f_{-k} f_{-l}
&= 2\,\I\, (g_n g_k g_l +h_n h_k g _h + h_n g_k h_l -g_n h_k h_l).\\
\end{split}
\end{equation}
Hence in the difference of nucleon propagators with opposite parity contains no terms proportional to $hhh$  as indicated in Fig. \ref{fig:nucgh}.

\section{$C(N_{\pm}^{(1)}) - C(N_{\pm}^{(2)})$ and $C(N_{\pm}^{(3)}) - C(N_{\pm}^{(4)})$}\label{app:C}

In order to prove that also the propagators $C(N_{\pm}^{(i)})$ and $C(N_{\pm}^{(i+1)})$ for $i=1,3$ are \textit{$g$-equivalent} we notice that from Table \ref{tab:BI} we have $\Gamma_2^{(i+1)} = \gamma_5 \Gamma_2^{(i)}$ and $\Gamma_1^{(i+1)} = \gamma_5 \Gamma_1^{(i)}$. 
Therefore from Eq. (\ref{eq:corr_expa}) and considering $i=1,3$, we have 
\begin{equation}
 \begin{split}
&\hspace*{-15mm}C(N_{\pm}^{(i)}) - C(N_{\pm}^{(i+1)}) =\pm \epsilon_{abc} \epsilon_{a'b'c'} (\mathcal{P}_{\pm})_{\alpha' \alpha} (\Gamma_2^{(i)})_{\beta\gamma}(\Gamma_2^{(i)})_{\gamma'\beta'}\\
&\hspace*{-15mm} \sum_{n>0, k>0, l>0} \Big[(f_n f_k f_l +f_{n}f_{-k} f_{-l}-f_{-n}f_k f_{-l}-f_{-n}f_{-k} f_{l})\\
 &\left[ \psi^{(n)}_{xa\alpha} \psi^{(n)\;\dagger}_{ya'\alpha'}
\psi^{(k)}_{xb\beta} \psi^{(k)\;\dagger}_{yb'\beta'}
\psi^{(l)}_{xc\gamma} \psi^{(l)\;\dagger}_{yc'\gamma'}- (\gamma_5)_{\alpha\xi}\psi^{(n)}_{xa\xi} \psi^{(n)\;\dagger}_{ya'\xi'}(\gamma_5)_{\xi'\alpha'}
\psi^{(k)}_{xb\beta} \psi^{(k)\;\dagger}_{yb'\beta'}(\gamma_5)_{\gamma\theta}\psi^{(l)}_{xc\theta} \psi^{(l)\;\dagger}_{yc'\theta'}(\gamma_5)_{\theta'\gamma'}\right]\\
&+(f_n f_k f_{-l} +f_{n}f_{-k} f_{l}-f_{-n}f_k f_{l}-f_{-n}f_{-k} f_{-l})\\
&\left[
\psi^{(n)}_{xa\alpha} \psi^{(n)\;\dagger}_{ya'\alpha'}
\psi^{(k)}_{xb\beta} \psi^{(k)\;\dagger}_{yb'\beta'}
(\gamma_5)_{\gamma\theta}\psi^{(l)}_{xc\theta} \psi^{(l)\;\dagger}_{yc'\theta'}
(\gamma_5)_{\theta'\gamma'}\right.
\left. -(\gamma_5)_{\alpha\xi}\psi^{(n)}_{xa\xi} \psi^{(n)\;\dagger}_{ya'\xi'}
(\gamma_5)_{\xi'\alpha'}
\psi^{(k)}_{xb\beta} \psi^{(k)\;\dagger}_{yb'\beta'}
\psi^{(l)}_{xc\gamma} \psi^{(l)\;\dagger}_{yc'\gamma'}
\right]\\
&-(f_n f_k f_l - f_{-n}f_k f_{-l})\\
&\left[ 
\psi^{(n)}_{xa\alpha} \psi^{(n)\;\dagger}_{yc'\gamma'}
\psi^{(k)}_{xb\beta} \psi^{(k)\;\dagger}_{yb'\beta'}
\psi^{(l)}_{xc\gamma} \psi^{(l)\;\dagger}_{ya'\alpha'}
-
(\gamma_5)_{\alpha\xi}\psi^{(n)}_{xa\xi} \psi^{(n)\;\dagger}_{yc'\xi'}(\gamma_5)_{\xi'\gamma'}
\psi^{(k)}_{xb\beta} \psi^{(k)\;\dagger}_{yb'\beta'}(\gamma_5)_{\gamma\theta}\psi^{(l)}_{xc\theta} \psi^{(l)\;\dagger}_{ya'\theta'}(\gamma_5)_{\theta'\alpha'}
\right]\\
&-(f_n f_k f_{-l} - f_{-n}f_k f_{l})\\
&\left[
\psi^{(n)}_{xa\alpha} \psi^{(n)\;\dagger}_{yc'\gamma'}
\psi^{(k)}_{xb\beta} \psi^{(k)\;\dagger}_{yb'\beta'}
(\gamma_5)_{\gamma\theta}\psi^{(l)}_{xc\theta} \psi^{(l)\;\dagger}_{ya'\theta'}(\gamma_5)_{\theta'\alpha'}\right.
\left.-
(\gamma_5)_{\alpha\xi}\psi^{(n)}_{xa\xi} \psi^{(n)\;\dagger}_{yc'\xi'}(\gamma_5)_{\xi'\gamma'}
\psi^{(k)}_{xb\beta} \psi^{(k)\;\dagger}_{yb'\beta'}
\psi^{(l)}_{xc\gamma} \psi^{(l)\;\dagger}_{ya'\alpha'}
\right]\\
&-(f_n f_{-k} f_{l} - f_{-n}f_{-k} f_{-l})\\
&\left[
\psi^{(n)}_{xa\alpha} \psi^{(n)\;\dagger}_{yc'\gamma'}
(\gamma_5)_{\beta\omega}\psi^{(k)}_{xb\omega} \psi^{(k)\;\dagger}_{yb'\omega'}(\gamma_5)_{\omega'\beta'}
\psi^{(l)}_{xc\gamma} \psi^{(l)\;\dagger}_{ya'\alpha'}\right.\\&
\left.-
(\gamma_5)_{\alpha\xi}\psi^{(n)}_{xa\xi} \psi^{(n)\;\dagger}_{yc'\xi'}(\gamma_5)_{\xi'\gamma'}
(\gamma_5)_{\beta\omega}\psi^{(k)}_{xb\omega} \psi^{(k)\;\dagger}_{yb'\omega'}(\gamma_5)_{\omega'\beta'}
(\gamma_5)_{\gamma\theta}\psi^{(l)}_{xc\theta} \psi^{(l)\;\dagger}_{ya'\theta'}(\gamma_5)_{\theta'\alpha'}
\right]\\
&-(f_n f_{-k} f_{-l} - f_{-n}f_{-k} f_{l})\\
&\left[
\psi^{(n)}_{xa\alpha} \psi^{(n)\;\dagger}_{yc'\gamma'}
(\gamma_5)_{\beta\omega}\psi^{(k)}_{xb\omega} \psi^{(k)\;\dagger}_{yb'\omega'}(\gamma_5)_{\omega'\beta'}(\gamma_5)_{\gamma\theta}\psi^{(l)}_{xc\theta} \psi^{(l)\;\dagger}_{ya'\theta'}(\gamma_5)_{\theta'\alpha'}
\right.\\&\left.-
(\gamma_5)_{\alpha\xi}\psi^{(n)}_{xa\xi} \psi^{(n)\;\dagger}_{yc'\xi'}(\gamma_5)_{\xi'\gamma'}
(\gamma_5)_{\beta\omega}\psi^{(k)}_{xb\omega} \psi^{(k)\;\dagger}_{yb'\omega'}(\gamma_5)_{\omega'\beta'}
\psi^{(l)}_{xc\gamma} \psi^{(l)\;\dagger}_{ya'\alpha'}
\right]\Big]\,.
 \end{split}
 \label{eq:c1c2}
 \end{equation}
Using (\ref{eq:f}) we can rewrite the coefficients in front of the eigenfunction products in terms of $h_n$ and $g_n$, namely
\begin{equation}
\begin{split}
f_n f_k f_l - f_{-n}f_k f_{-l}&=-2(h_n g_k g_l +\I\, h_n h_k g_l +g_n g_k h_l +\I\, g_n h_k h_l),\\
f_n f_k f_{-l} - f_{-n}f_k f_l&=-2(h_n g_k g_l +\I\, h_n h_k g_l -g_n g_k h_l -\I\, g_n h_k h_l),\\
f_n f_{-k} f_l - f_{-n}f_{-k} f_{-l}&=-2(h_n g_k g_l -\I\, h_n h_k g_l +g_n g_k h_l -\I\, g_n h_k h_l),\\
f_n f_{-k} f_{-l} - f_{-n}f_{-k} f_l&=-2(h_n g_k g_l -\I\, h_n h_k g_l -g_n g_k h_l +\I\, g_n h_k h_l).
\end{split}
\label{eq:fff2}
\end{equation}
Therefore as we can see from Eqs. (\ref{eq:c1c2}) and (\ref{eq:fff2}) the differences $C(N_{\pm}^{(1)}) - C(N_{\pm}^{(2)})$ and $C(N_{\pm}^{(3)}) - C(N_{\pm}^{(4)})$ are only proportional to $ggh$ and $ghh$ terms, as indicated in Fig. \ref{fig:nucgh}.
%


\end{document}